\documentclass[preprintnumbers,amsmath,
amssymb]{revtex4}

\usepackage[english]{babel}

\usepackage[letterpaper,top=2cm,bottom=2cm,left=3cm,right=3cm,marginparwidth=1.75cm]{geometry}

\usepackage{amsmath}
\usepackage{amssymb}
\usepackage{subcaption}
\usepackage{graphicx}
\usepackage[colorlinks=true, allcolors=blue]{hyperref}
 
\begin{document}

%
\title{Space-resolved 
Chemical Information 
from Infrared 
Extinction 
Spectra 
} 
\author{Kalpa de Silva$ ^1$, 
Proity Akbar$ ^2$, and Reinhold 
Bl\"umel$ ^1$} 
\affiliation{$^1$Department of Physics, Wesleyan University, 
Middletown, Connecticut 06459-0155}
\affiliation{$^2$Department of Chemistry, Wesleyan University, 
Middletown, Connecticut 06459-0155}

\begin{abstract}
Based on the 
classical Lorentz model 
of the index of refraction,
a new method  
is presented for the 
extraction of 
the complex index of refraction from 
the extinction efficiency of 
homogeneous and layered 
dielectric spheres that 
simultaneously removes 
scattering effects and 
corrects measured 
extinction spectra for 
baseline shifts, tilts, 
curvature, and 
scaling. No reference spectrum 
is required and the method 
automatically satisfies 
the Kramers-Kronig relations. 
Thus, the method 
yields  
pure absorbance spectra for 
unambiguous interpretation 
of the chemical 
information of the sample. 
In the case of homogeneous 
spheres, the method also determines 
the radius of the sphere. 
In the case of layered spheres, 
the method determines the 
radii of the layers 
and the 
substances within each layer. 
Only a single-element detector 
is required. 
Using  
simulated $Q_{ext}$ 
data of 
polymethyl-methacrylate 
(PMMA) and polystyrene 
homogeneous and 
layered spheres 
we show 
that our reconstruction 
algorithm is reliable and 
accurately extracts pure 
absorbance spectra from 
$Q_{ext}$ data. 
Reconstructing the 
pure absorbance spectrum 
from a published, experimentally 
measured raw absorbance 
spectrum shows that our 
method also works for 
uncorrected spectra, 
reliably removes 
scattering effects and, 
given shape information, 
simultaneously corrects 
for spectral distortions. 
\end{abstract}
 
 
\maketitle
 
\section{Introduction} 
\label{INTRO}
With roots going back 
several decades in time,
infrared (IR) 
spectroscopy 
has evolved into one of 
the most powerful 
and successful 
tools for the 
analysis 
of biological cells 
and tissues 
for applications in 
biophysics, 
chemistry, 
and medical pathology
\cite{review-1,MOH,
LUK1,SC-1}. 
Far from being a 
successful, but stagnant 
field of study, 
with the recent advent 
of tunable IR laser 
sources 
\cite{review-2}, 
IR spectroscopy 
is currently experiencing 
growing interest and 
expansion into new 
fields of application. 
No matter whether 
the spectra are taken 
with thermal, 
laser, 
or synchrotron IR 
sources, 
two tasks always 
have to be performed: 
(1) Cleaning up 
of the raw spectra and 
(2) reconstructing 
the refractive 
index (revealing 
the chemistry) of the 
(biological) 
sample under 
investigation. 
Powerful tools exist 
for cleaning up the 
spectra, 
i.e., 
removing artefacts 
introduced by the 
spectrographs and 
IR sources that result, 
e.g., in baseline shifts,  
tilts, 
curvature, 
and multiplicative 
distortions of 
the spectra 
\cite{ME-EMSC,van2013recovery}. 
 
This paper focuses 
on single-cell 
IR spectroscopy where 
the cell sizes are 
approximately 
of the same order of 
magnitude 
as the wavelength of 
the incident IR radiation. 
Thus, the corresponding IR spectra 
exhibit pronounced 
scattering effects, 
such as Mie scattering 
\cite{van-de-hulst}, 
that modify absorbance 
spectra considerably 
and need to be removed 
before interpreting 
the spectra. 
 
In this paper we present 
a new technique that 
is based on the classical 
Lorentz model of the 
index of refraction 
\cite{Griffiths} and 
is capable of accomplishing 
the tasks (1) and (2) 
above without the 
need of a reference 
spectrum 
\cite{ME-EMSC}
or the Kramers-Kronig 
relations 
\cite{ME-EMSC,van2013recovery}
as required 
by leading 
spectral correction 
methods. 
For our examples, we 
use two substances, 
polymethyl-methacrylate 
(PMMA) and 
polystyrene
(PS). 
%
%

\section{Methods} 
\label{METH}
In subsection 
\ref{METH-GEN}
we start  
by describing 
the general method 
used in IR spectroscopy.
Then, in subsection 
\ref{METH-LORENTZ},
we describe 
our method of 
reconstructing 
the space- and 
wavenumber-dependent 
complex index of 
refraction 
$\eta(\vec r,\tilde\nu)$, 
i.e., the hyperspectral 
complex refractive index, 
that is based on 
the classical Lorentz  
model 
\cite{Griffiths}
of the index 
of refraction. 
Concluding this 
section, 
based on the 
methods described 
in 
subsections 
\ref{METH-GEN} 
and 
\ref{METH-LORENTZ},
we describe in 
subsection
\ref{METH-SIMUL} 
our numerical procedures  
for reconstructing 
the complex refractive 
index 
$\eta(\vec r,\tilde\nu)$, 
and with it 
pure absorbance spectra 
(i.e., $\Im[
\eta(\vec r,\tilde\nu)]$). 
\subsection{General}
\label{METH-GEN}
Typically in  
single-detector 
IR spectroscopy,  
a beam of 
IR radiation 
of wavelength $\lambda$ 
(wavenumber 
$\tilde \nu=1/\lambda$)
and intensity 
$I_0(\tilde\nu)$
is directed toward a 
target that may be 
a biological cell, 
a sample of biological 
tissue, or a grain 
of biological 
or inanimate matter, 
and the transmitted 
intensity $I(\tilde\nu)$ 
is measured 
in forward direction. 
Because of absorption 
of IR radiation by 
the target, and 
because of scattering 
of IR radiation 
out of the 
forward direction, 
we have 
$I(\tilde\nu)
<I_0(\tilde\nu)$. 
Once $I(\tilde\nu)$ 
is measured, the 
result of the measurements 
is typically 
reported as 
the apparent 
absorbance 
\begin{equation}
{\cal A}(\nu) = 
-\log_{10}
\left[\frac{I(\tilde\nu)}
{I_0(\tilde\nu)}
\right] .
\label{METH-1}
\end{equation}

While the apparent 
absorbance 
${\cal A}$ 
is convenient for 
recording and 
reporting experimental 
results, 
cross sections are 
more convenient 
for analysis of 
experiments,  
numerical 
simulations, and the 
extraction of 
refractive 
indexes. 
We denote by 
$\sigma_{\rm scat}$ 
and 
$\sigma_{\rm abs}$
the scattering 
and absorption 
cross sections, 
respectively. 
Then, the 
extinction 
cross section 
is defined as 
\cite{van-de-hulst} 
\begin{equation}
\sigma_{\rm ext} = 
\sigma_{\rm scat} + 
\sigma_{\rm abs} .
\label{METH-2}
\end{equation}
We also define 
scattering, 
absorption, 
and 
extinction 
efficiencies 
according to 
\cite{van-de-hulst}
\begin{equation}
Q_{\rm scat} = 
\sigma_{\rm scat}/g,
\ \ \ 
Q_{\rm abs} = 
\sigma_{\rm abs}/g,
\ \ \ 
Q_{\rm ext} = 
\sigma_{\rm ext}/g,
\label{METH-3}
\end{equation}
respectively, 
where $g$ is the 
geometric 
cross section of 
the sample under 
investigation. 
Denoting by 
$G\gg g$ the 
reception area 
of the detector, 
the radiative power 
received by the 
detector without 
the sample present 
is 
\begin{equation}
P_0 = G I_0
\label{METH-4}
\end{equation}
and the radiative 
power received by 
the detector 
with the sample 
present is 
\begin{equation}
P_I=GI=
GI_0-I_0\sigma_{ext}. 
\label{METH-5}
\end{equation}
From this, we obtain 
\begin{equation}
Q_{\rm ext} = 
\frac{\sigma_{\rm ext}}
{g} = 
\frac{GI_0-GI}{gI_0}
=\left(\frac{G}{g}\right)
\left[1-\left(\frac{I}{I_0}
\right)\right] . 
\label{METH-6}
\end{equation}
Via (\ref{METH-1}), 
this can immediately be 
related to 
the apparent 
absorbance 
${\cal A}$ 
according to 
\begin{equation}
Q_{\rm ext} = 
\left(\frac{G}{g}\right) 
\left[ 1-10^{-{\cal A}}
\right] .
\label{METH-7}
\end{equation}
Since, via 
(\ref{METH-7}), 
$Q_{\rm ext}$ and 
${\cal A}$ 
contain the same 
information, 
we will, in the following, 
present our 
theoretical analysis 
and reconstructions 
of refractive indexes  
on the 
basis of $Q_{\rm ext}$ 
or ${\cal A}$ depending 
on convenience. 
 
%
%
 
\subsection{Reconstruction 
via Classical Lorentz Model}
\label{METH-LORENTZ}
The main point of our 
method is to extract 
the space- and 
wavenumber-dependent 
complex index of 
refraction  
$\eta(\vec r,\tilde\nu)$ 
from 
$Q_{\rm ext}(\tilde \nu)$. 
This may sound 
ambitious, since 
$Q_{\rm ext}(\tilde \nu)$ 
has no space dependence. 
Nevertheless, 
it is possible, at least 
for a sphere with 
two layers, 
as demonstrated in 
Section~\ref{RESULTS}. 
To implement the method, 
we discretize space 
into voxels, 
which can be any 
shape or size. 
The smaller, the 
more resolution. 
The more regular 
(such as cubes or shells), 
the easier the 
computations. 
Labeling the 
voxels with the 
discrete index 
$j=1,\ldots,J$, 
and denoting 
by $\vec r_j$ 
a representative 
point of voxel 
number $j$ 
(for instance its 
center in the case 
of cubes or a point 
on the center shell 
in the case of 
shells), 
we represent 
the refractive index 
inside of a voxel 
by its (spatially)
constant value 
$\eta(\vec r_j,\tilde\nu)
\equiv \eta_j(\tilde\nu)$, 
which varies in wavenumber  
as a sum of 
Lorentzians of the 
form 
\begin{equation}
\eta_j(\tilde\nu) = 
\sum_{m=1}^M 
\frac{h_m^{(j)}}
{1+[(\tilde\nu-
\tilde\nu_m^{(j)}/
\Gamma_m^{(j)}]^2} , 
\ \ \ j=1,\ldots,J , 
\label{L-sum}
\end{equation}
where, 
inside voxel 
number $j$, 
$\tilde\nu_m^{(j)}$ is 
the wavenumber location 
of an IR absorption band,
$h_m^{(j)}$ is the 
corresponding peak height, 
$\Gamma_m^{(j)}$ 
is the 
width of the absorption 
band, 
$m=1,\ldots,M$ 
numbers the absorption 
bands, and $M$, 
a parameter, 
decides how many bands 
we include in our 
reconstructions. 
$\tilde\nu_m^{(j)}$,
$h_m^{(j)}$, and 
$\Gamma_m^{(j)}$ 
are adjustable parameters 
to be determined 
via a 3D scattering code 
such 
that the 
$Q_{\rm ext}(\tilde\nu)$ 
predicted 
on the basis of 
(\ref{L-sum}) 
optimally reproduces 
$Q_{\rm ext}^{({\rm given})}
(\tilde\nu)$, 
where 
$Q_{\rm ext}^{({\rm given})}
(\tilde\nu)$ 
is the given 
$Q_{\rm ext}(\tilde\nu)$, 
either obtained experimentally, 
or from a 
numerical simulation 
(see Section~\ref{RESULTS}). 

In the simplest case, 
the entire scatterer may 
be represented as a single 
voxel, in which case $M=1$. 
This is appropriate for 
homogeneous scatterers, i.e., 
scatterers with a spatially 
constant index of refraction, 
such as a homogeneous sphere. 
Another example is a  
sphere with two layers, 
i.e., a core and a shell,
each homogeneously filled 
with a medium of spatially 
constant index of refraction
as discussed 
in Section~\ref{RESULTS}. 
In this case two voxels, 
i.e., the core 
and the shell, 
suffice, each 
endowed with its 
own series (\ref{L-sum}) 
of Lorentzians. 
The method now determines 
the 
unknown parameters 
$\tilde\nu_m^{(j)}$,
$h_m^{(j)}$,
and $\Gamma_m^{(j)}$ 
as the optimum values 
that best reproduce 
$Q_{\rm ext}
^{({\rm given})}
(\tilde\nu)$. 
This can be done with any 
standard nonlinear 
fit algorithm that 
minimizes a target function 
$S$, 
for instance, 
\begin{equation}
S[\tilde\nu_m^{(j)},
h_m^{(j)},
\Gamma_m^{(j)}] = 
\sum_{l} \left\{ 
Q_{\rm ext}
^{({\rm model})} 
[\tilde\nu_m^{(j)},
h_m^{(j)},
\Gamma_m^{(j)};\tilde\nu_l]
-
Q_{\rm ext}
^{({\rm given})}
(\tilde\nu_l)
\right\}^2, 
\label{target}
\end{equation}
where 
$Q_{\rm ext}
^{({\rm model})} 
[\tilde\nu_m^{(j)},
h_m^{(j)},
\Gamma_m^{(j)};\tilde\nu_l]$ is the 
extinction 
efficiency evaluated 
at $\tilde\nu_l$
with fit parameters 
$\tilde\nu_m^{(j)}$,
$h_m^{(j)}$,
and $\Gamma_m^{(j)}$, 
computed via a suitable 
3D scattering code
and 
$\tilde\nu_l$ are 
the wavenumbers at which 
$Q_{\rm ext}
^{({\rm given})}(\tilde\nu)$ 
has been 
measured or pre-determined 
numerically. 

The main characteristic 
of the Lorentzian 
method is that, 
in contrast to 
other spectral-correction 
methods (e.g., 
\cite{ME-EMSC}), 
no reference spectrum 
is needed to 
scatter-correct 
$Q_{\rm ext}$ spectra, 
whether simulated 
via a forward model 
or obtained
experimentally, 
and that no 
Kramers-Kronig 
transformation is 
needed. This is so, 
since the 
complex Lorentzian 
functions in 
(\ref{L-sum}) 
automatically 
satisfy the 
Kramers-Kronig 
relations, and 
thus $\eta_j(\tilde\nu)$, 
as a finite sum of 
such functions, 
automatically 
satisfies the 
Kramers-Kronig 
relations as well. 

%
 
\subsection{Numerical Simulations} 
\label{METH-SIMUL}
To demonstrate the 
power of the 
Lorentzian method, 
we performed 
four types of 
numerical simulations, 
labeled I, II, 
III, and IV. 
Types I, II, and III 
are for homogeneous spheres 
and type IV is for a 
layered sphere. 
 
In our type-I simulations, 
we fit 
accurate experimental 
data for $\eta(\tilde\nu)$ of 
PMMA 
\cite{PMMA-refr-idx}
and PS 
\cite{PS-refr-idx} 
with  
$M=18$ (PMMA) and 
$M=22$ (PS) Lorentzians, 
respectively, and compute 
$Q_{\rm ext}^{({\rm given})}
(\tilde\nu)$ 
with 
$\eta$ in (\ref{L-sum}) 
using a standard Mie 
scattering code. 
The results of these 
type-I simulations are 
presented and discussed 
in Section~\ref{METH-LQEX}. 
 
In our type-II simulations, 
we compute 
$Q_{\rm ext}^{({\rm given})}
(\tilde\nu)$ 
directly on the basis 
of the experimental 
data for $\eta$. 
The  
results 
of our  
type-II simulations are 
presented and discussed 
in Section~\ref{ELLQEX}. 
 
Our type-III simulations 
take 
$Q_{\rm ext}^{({\rm given})}
(\tilde\nu)$ 
directly from an experimental 
absorbance spectrum 
published in 
\cite{blumel2016PMMA}. 
In this case we also 
perform baseline, tilt, 
scaling, and curvature 
corrections of the 
experimental raw spectrum, 
by adding the baseline-, 
tilt-, scaling-, and 
curvature parameters 
as additional fit parameters 
to our Lorentzian method. 
As shown in 
Section~\ref{EXPQEX} 
this works very well, 
and thus our Lorentzian 
method performs both 
the correction of 
raw spectra and 
scattering corrections 
in a single step, 
without the need of 
a reference spectrum 
or the Kramers-Kronig 
transform. 
 
In our type-IV simulations, 
we use our Lorentzian method 
to reconstruct the 
complex 
space-dependent index of 
refraction for a sphere 
with two layers, i.e., 
PMMA in the core and PS 
in the shell. 
This is a case with two 
voxels, i.e., 
$j=1$ representing the 
complex refractive index 
in the core and 
$j=2$ representing the 
complex refractive index 
in the shell. 
In this case, 
we construct 
$Q_{\rm ext}^{({\rm given})}
(\tilde\nu)$ 
using a standard Mie code 
for stratified spheres 
with 
$\eta^{({\rm PMMA})}
(\tilde\nu)$ 
in the core and 
$\eta^{({\rm PS})}
(\tilde\nu)$ 
in the shell, both 
taken directly from 
the corresponding 
experimental 
complex 
refractive 
indexes. 
The results of 
these simulations, 
reported in 
Section~\ref{LAY-SPH},
prove that the Lorentzian 
method with only a single-element 
detector, and a single 
$Q_{\rm ext}(\tilde\nu)$ 
as input, can reveal 
space-dependent chemistry, 
at least for this two-layer 
model.  
 
One characteristic feature 
of our Lorentzian method 
is that, initially, 
as starting conditions, 
we place $M$ Lorentzians, 
equi-spaced (unbiased)
into the 
wavenumber interval 
under consideration. 
This way, in addition to 
using random initial values for 
Lorentzian
peak heights and peak 
widths, we obtain an 
initial condition that 
is not only unbiased, 
but an initial Lorentzian 
peak is always near to 
an actual absorption peak 
in the pure absorbance 
spectrum, which then needs 
only a relatively small 
adjustment in position, 
height, and widths, to 
morph into the actual 
absorption peak. 
This feature is crucial. 
It avoids, to a large extent, 
getting trapped in 
secondary optimization 
minima, i.e., 
a stationary but not 
global minimum of 
the target function $S$  
[see (\ref{target})].
We can still get trapped 
in secondary minima, but 
because of our choice 
of equi-spaced 
initial Lorentzian positions, 
the chance of getting trapped 
in a secondary minimum that 
results in a particularly 
bad fit is vastly reduced. 

%
%
 
\section{Results} 
\label{RESULTS} 
In this section we apply 
our Lorentzian reconstruction 
method to two important cases: 
(a) Homogeneous spheres 
(see Section~\ref{HOM-SPH}) 
and (b) layered spheres 
(see Section~\ref{LAY-SPH}). 
While these examples are 
important for testing 
validity and robustness 
of our method, IR 
spectroscopy on homogeneous 
and layered spheres may also 
have practical importance, 
for instance in 
mineralogy, or for 
characterizing 
micro- and nano-spheres, 
such as 
PMMA  
or 
PS spheres. 
PMMA spheres, 
in particular, are 
frequently used as 
model systems for 
biological cells 
\cite{van2013recovery,
blumel2016PMMA,LUK1,
bassan2009resonant}. 
 
%
%
\subsection{Homogeneous Spheres 
(types I, II, and III)} 
\label{HOM-SPH} 
In this section  
we demonstrate 
the validity of our 
reconstruction method 
for homogeneous spheres, 
all of radius 
$10\mu$m, 
in three steps, working 
with two different 
substances: 
PMMA and PS. 
We start 
by simulating 
PMMA and PS 
$Q_{\rm ext}^{({\rm given})}
(\tilde\nu)$
spectra based on a 
Lorentzian approximation 
of the complex refractive 
index 
and show that for all 
three substances 
our method exactly 
reconstructs the complex 
index of refraction 
(see Section~\ref{METH-LQEX}). 
Then, we simulate 
PMMA and PS 
$Q_{\rm ext}^{({\rm given})}
(\tilde\nu)$
spectra 
based directly 
on experimental  
data of the complex index 
of refraction and demonstrate 
that in this case, as well, 
near-perfect reconstructions 
are obtained 
(see Section~\ref{ELLQEX}). 
Finally, we reconstruct 
the complex index of refraction 
from an experimentally measured 
PMMA spectrum 
\cite{blumel2016PMMA}, 
which also includes 
baseline, tilt,  
scale, and 
curvature corrections  
(see Section~\ref{EXPQEX}).
We show that 
a satisfactory result 
is achieved in 
this case as well. 
%
%
\subsubsection{Lorentzian 
Extinction (type-I)}
\label{METH-LQEX} 
\begin{figure}
\centering
\begin{subfigure}{0.45\textwidth}
    \includegraphics[width=\textwidth]
    {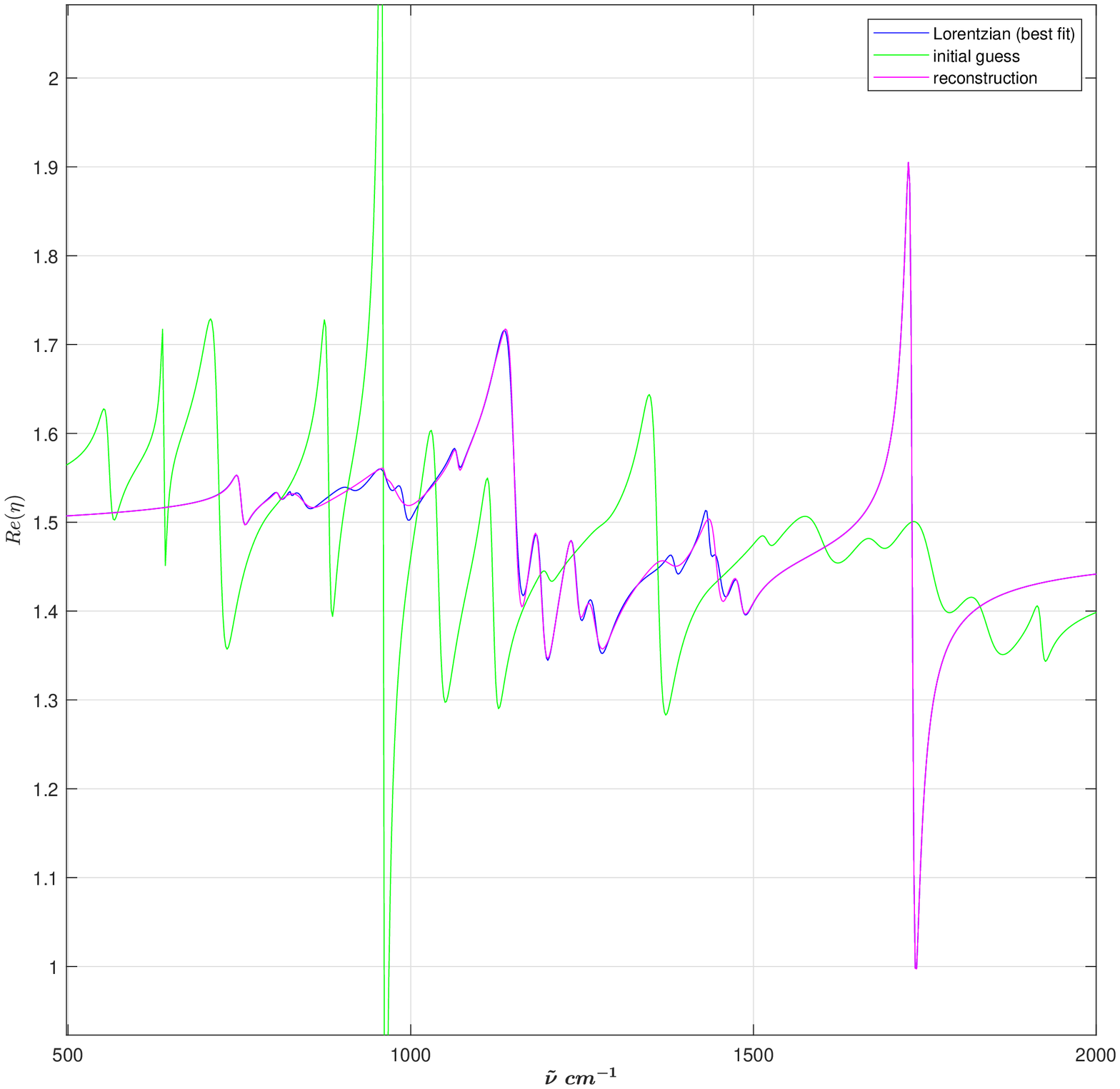}
    \caption{PMMA reconstruction; 
    real part of $\eta$.}
    \label{fig1a}
\end{subfigure}
\hfill
\begin{subfigure}{0.45\textwidth}
    \includegraphics[width=\textwidth]
    {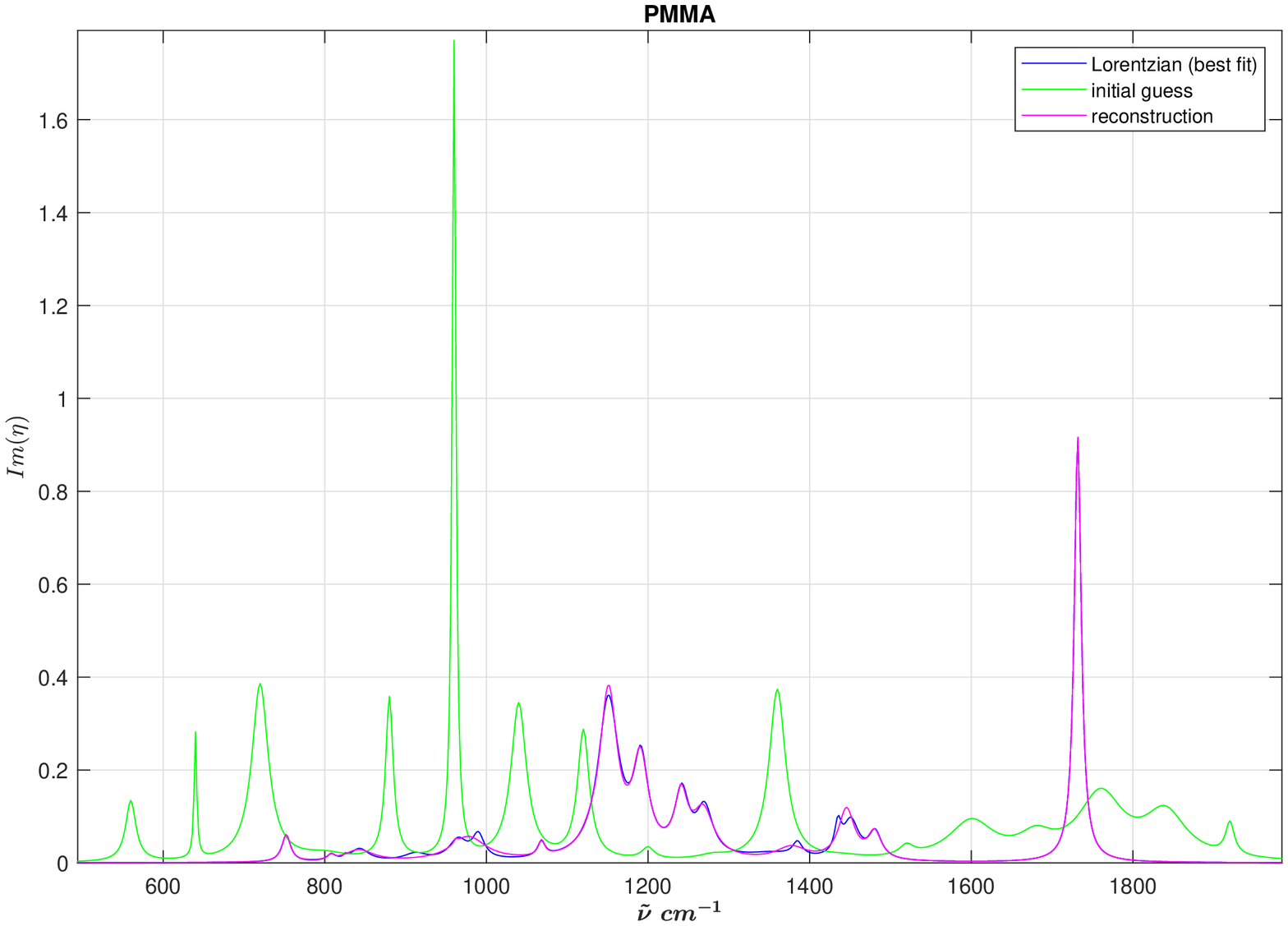}
    \caption{PMMA reconstruction; 
    imaginary part of $\eta$.}
    \label{fig1b}
\end{subfigure}
\hfill
\begin{subfigure}{0.45\textwidth}
    \includegraphics[width=\textwidth]
    {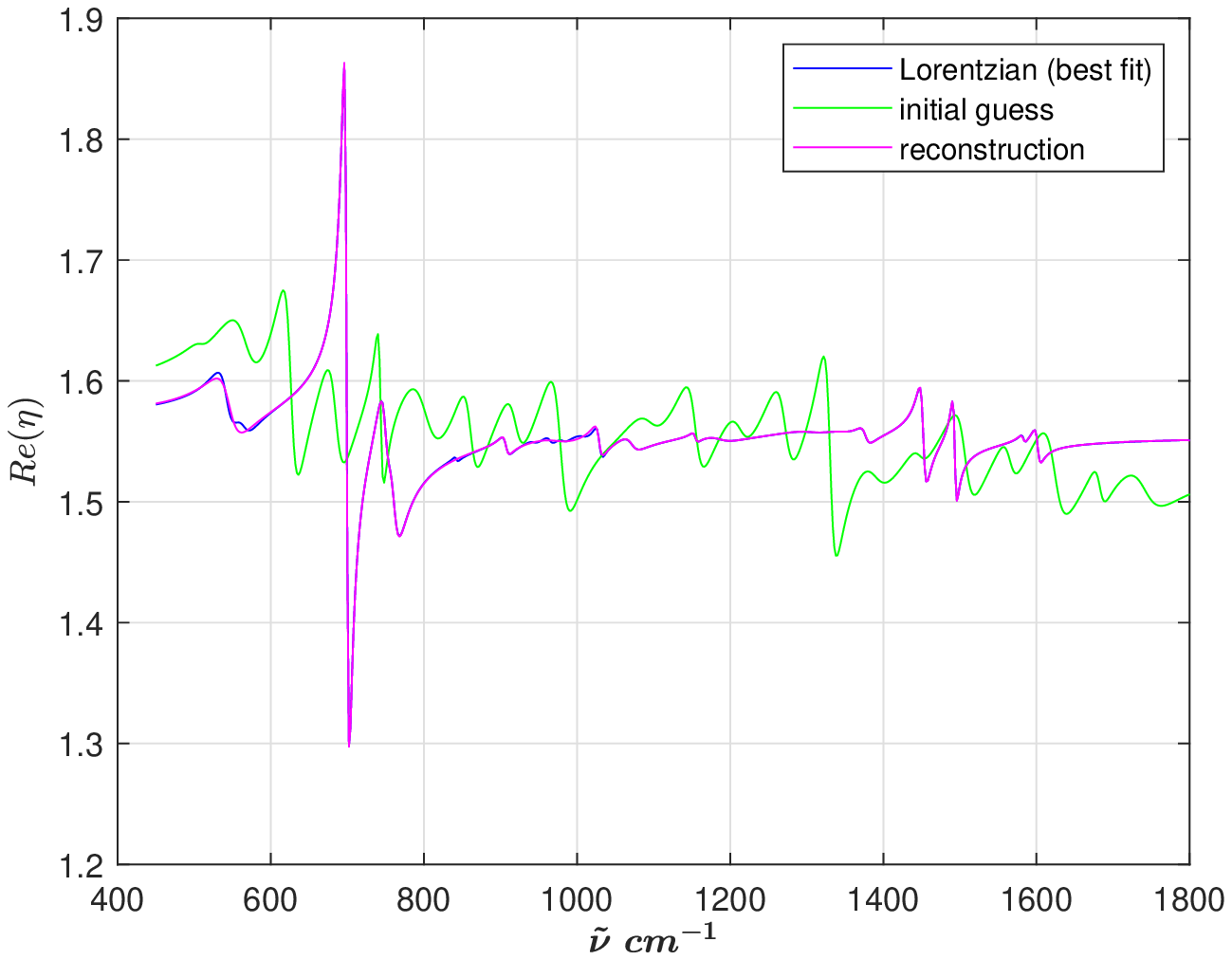}
    \caption{PS reconstruction; 
    real part of $\eta$.}
    \label{fig1c}
\end{subfigure}
\hfill
\begin{subfigure}{0.45\textwidth}
    \includegraphics[width=\textwidth]
    {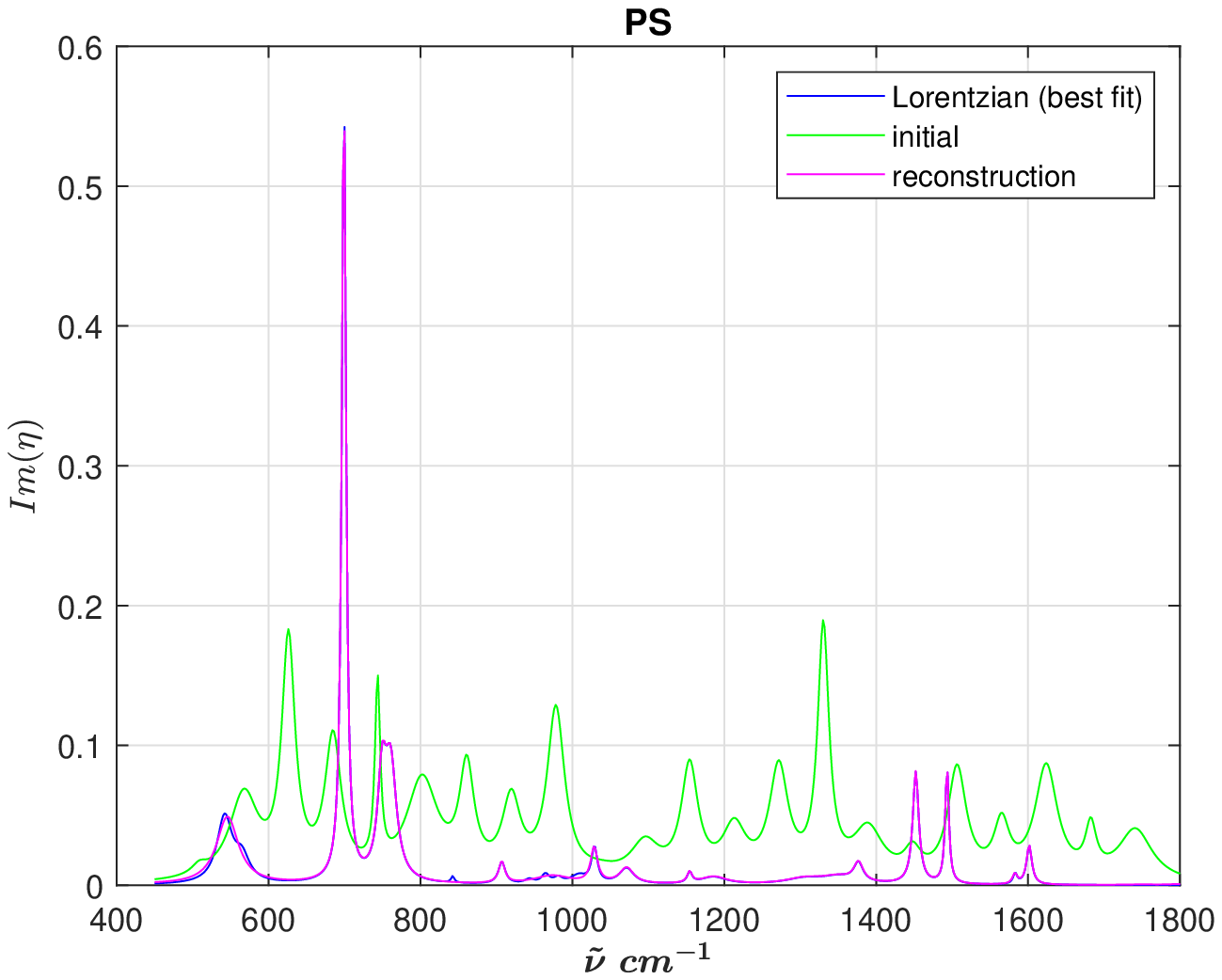}
    \caption{PS reconstruction; 
    imaginary part of $\eta$.}
    \label{fig1d}
\end{subfigure}
\caption{Reconstructions of 
real and imaginary 
parts of the 
complex refractive index $\eta$
of PMMA and PS using a Lorentzian 
fit to experimental data 
(blue lines) 
for 
obtaining 
$Q_{\rm ext}^{({\rm given})}
(\tilde\nu)$. 
18 Lorentzians were used for 
the PMMA reconstructions and 
22 Lorentzians were used for 
the PS reconstructions.
Although the initial guess 
of the complex refractive indexes 
is random (green lines), 
the reconstructions (magenta 
lines)
are near perfect.}
\label{fig:fig1}
\end{figure}

Approximating the experimental  
data of $\eta$ by 
a set of Lorentzians, 
and computing 
$Q_{\rm ext}^{({\rm given})}
(\tilde\nu)$, 
on the basis of these 
Lorentzians, 
we obtain the 
reconstructions of 
the real and imaginary 
parts of the 
refractive index 
as shown in 
Fig.~\ref{fig:fig1}. 
Although we start with 
initial conditions 
(green lines) 
that 
are far off the final 
result (magenta lines), 
we see that our 
reconstructions 
(magenta lines) 
overlap the target 
refractive indexes 
(blue lines) nearly perfectly. 
This shows that our 
type-I 
reconstructions are near perfect. 
Given that we start our 
reconstruction algorithm 
(green lines)
far from the target 
(blue lines) 
it also shows that our 
Lorentzian reconstruction 
algorithm 
is robust. 
Adding to the robustness, 
we also simulated 
reconstructions 
(not shown) adding 
10\% noise to 
$Q_{\rm ext}^{({\rm given})}
(\tilde\nu)$. 
We obtained reconstructed 
spectra of similar accuracy. 
This shows that our 
reconstruction algorithm 
is robust also to 
perturbations in 
$Q_{\rm ext}^{({\rm given})}
(\tilde\nu)$.
Thus, the
main result of this section 
is that our Lorentzian method 
works well and is robust 
with respect to initial 
conditions and noise. 

%
%

%
%
 
\subsubsection{Extinction 
based on experimental refractive 
index (type-II)} 
\label{ELLQEX}
\begin{figure}
\centering
\begin{subfigure}{0.45\textwidth}
    \includegraphics[width=\textwidth]
    {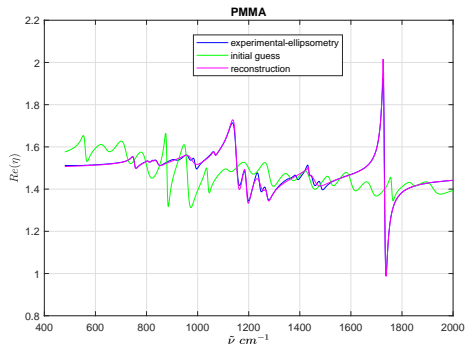}
    \caption{PMMA reconstruction; 
    real part of $\eta$.}
    \label{fig2a}
\end{subfigure}
\hfill
\begin{subfigure}{0.45\textwidth}
    \includegraphics[width=\textwidth]
    {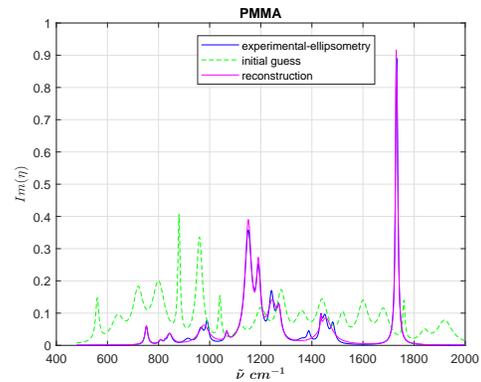}
    \caption{PMMA reconstruction; 
    imaginary part of $\eta$.}
    \label{fig2b}
\end{subfigure}
\hfill
\begin{subfigure}{0.45\textwidth}
    \includegraphics[width=\textwidth]
    {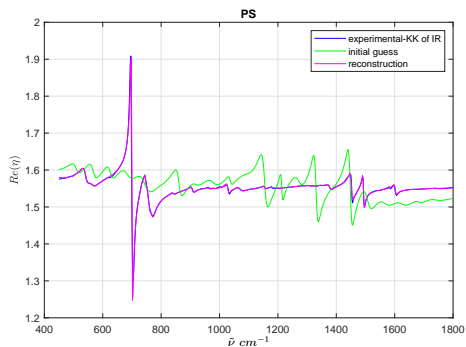}
    \caption{PS reconstruction; 
    real part of $\eta$.}
    \label{fig2c}
\end{subfigure}
\hfill
\begin{subfigure}{0.45\textwidth}
    \includegraphics[width=\textwidth]
    {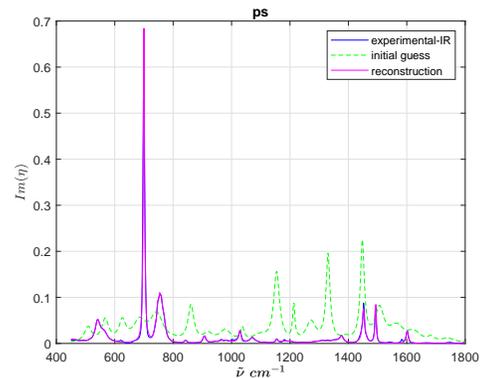}
    \caption{PS reconstruction; 
    imaginary part of $\eta$.}
    \label{fig2d}
\end{subfigure}
        
\caption{Reconstructions of 
real and imaginary parts of the 
complex refractive index $\eta$
of PMMA and PS using 
experimental data 
(blue lines) 
for 
obtaining 
$Q_{\rm ext}^{({\rm given})}
(\tilde\nu)$. 
18 Lorentzians were used for 
the PMMA reconstructions and 
22 Lorentzians were used for 
the PS reconstructions.
Although the initial guess 
of the complex refractive indexes 
is random (green lines), 
the reconstructions (magenta 
lines)
are near perfect.}
\label{fig:fig2}
\end{figure}

In this section we make it 
slightly harder for our 
reconstruction algorithm 
by not approximating the 
experimental refractive index 
by Lorentzians, but 
compute 
$Q_{\rm ext}^{({\rm given})}
(\tilde\nu)$ 
directly on the basis of 
the experimental complex  
$\eta$. 
The result is shown in 
Fig.~\ref{fig:fig2}. 
Again, although starting 
far from the optimal values 
of the Lorentzian fit parameters, 
we obtain near perfect 
reconstructions of real and 
imaginary refractive indexes. 
In this case, too, we checked 
for the influence of noise, 
and found our algorithm robust 
with respect to noise perturbations. 
%
 
%
%
 
\subsubsection{Experimental  
Extinction (type-III)}
\label{EXPQEX} 
\begin{figure}
\centering
\begin{subfigure}{0.45\textwidth}
    \includegraphics[width=\textwidth]
    {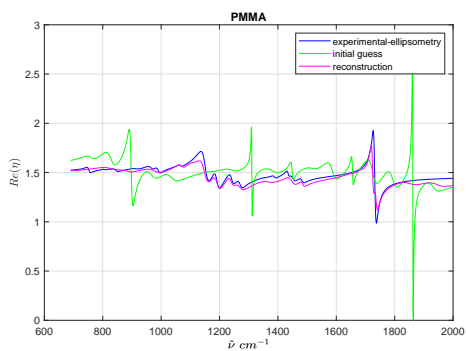}
    \caption{PMMA reconstruction 
    from experimental data; 
    real part of $\eta$.}
    \label{fig3a}
\end{subfigure}
\hfill
\begin{subfigure}{0.45\textwidth}
    \includegraphics[width=\textwidth]
    {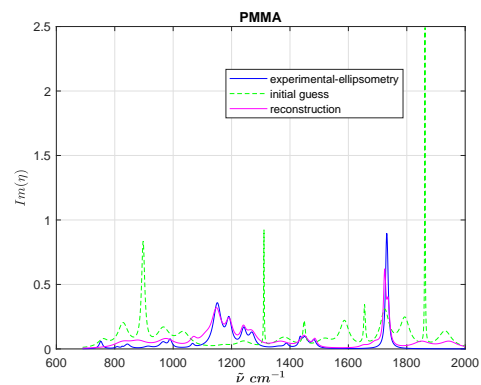}
    \caption{PMMA reconstruction 
    from experimental data;
    imaginary part of $\eta$.}
    \label{fig3b}
\end{subfigure}
        
\caption{Reconstructions of 
real and imaginary parts of the 
complex refractive index $\eta$
of PMMA 
(blue lines) 
for an experimental 
PMMA extinction 
spectrum of 
a PMMA sphere
\cite{blumel2016PMMA}. 
18 Lorentzians were used for 
the reconstructions.
Although the initial guess 
of the complex refractive index 
is random (green lines), 
the reconstructions (magenta 
lines) of real and imaginary 
parts are close to their 
experimental values.}
\label{fig:fig3}
\end{figure}

In our third numerical experiment 
concerned with homogeneous 
spheres, we reconstructed 
the complex PMMA refractive 
index from an experimental 
apparent absorbance spectrum 
\cite{blumel2016PMMA}, 
simultaneously performing 
baseline-, tilt-, 
scaling-, 
and curvature corrections 
of the the raw spectrum. 
The result is shown in 
Fig.~\ref{fig:fig3}. 
Starting our Lorentzian 
reconstruction algorithm 
once more with random 
initial conditions, 
we obtained reconstructions 
(magenta lines) 
of the real and imaginary 
parts of the complex PMMA 
index of refraction in 
satisfactory agreement 
with the experimental 
data 
(blue lines). 
This shows that our algorithm 
also works for experimental 
raw spectra that are neither 
scatter corrected, nor 
corrected for experimental 
systematic errors. 
Our algorithm can handle 
both simultaneously without 
the need of a reference 
spectrum or the 
Kramers-Kronig transformation. 
 
%
%

 
\subsection{Layered Sphere 
(type-IV)} 
\label{LAY-SPH} 
\begin{figure}
\centering
\begin{subfigure}{0.4\textwidth}
    \includegraphics[width=\textwidth]
    {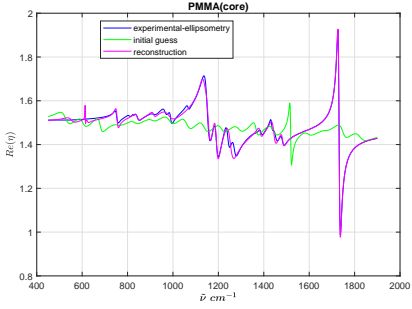}
    \caption{Reconstruction 
    of the real part 
    of $\eta$ of the core 
    material; it is correctly 
    identified as the real part 
    of $\eta$ of PMMA.}
    \label{fig4a}
\end{subfigure}
\hfill
\begin{subfigure}{0.4\textwidth}
    \includegraphics[width=\textwidth]
    {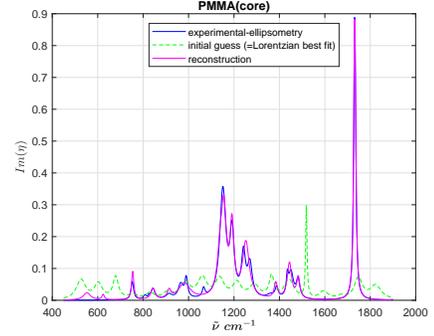}
    \caption{Reconstruction 
    of the imaginary part 
    of $\eta$ of the core 
    material; it is correctly 
    identified as the 
    imaginary part 
    of $\eta$ of PMMA.}
    \label{fig4b}
\end{subfigure}
\hfill
\begin{subfigure}{0.4\textwidth}
    \includegraphics[width=\textwidth]
    {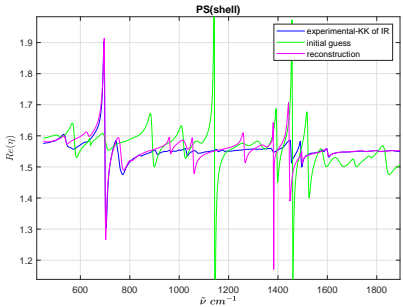}
    \caption{Reconstruction 
    of the real part 
    of $\eta$ of the shell 
    material; it is correctly 
    identified as the real part 
    of $\eta$ of PS.}
    \label{fig4c}
\end{subfigure}
\hfill
\begin{subfigure}{0.4\textwidth}
    \includegraphics[width=\textwidth]
    {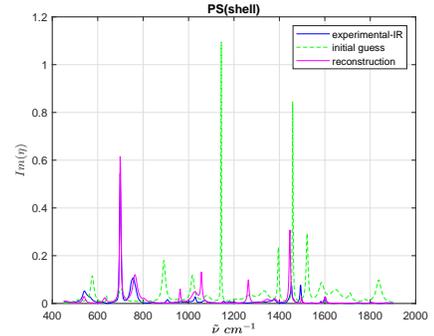}
    \caption{Reconstruction 
    of the imaginary part 
    of $\eta$ of the shell
    material; it is correctly 
    identified as the 
    imaginary part 
    of $\eta$ of PS.}
    \label{fig4d}
\end{subfigure}
\caption{Reconstructions of 
real and imaginary 
parts of the 
complex refractive 
index $\eta$
(blue lines) 
for a layered sphere 
consisting of PMMA in the core 
and PS in the shell. 
$Q_{\rm ext}^{({\rm given})}
(\tilde\nu)$
was produced via a standard 
Mie scattering code 
for stratified spheres. 
18 Lorentzians were used for 
the PMMA reconstructions and 
22 Lorentzians were used for 
the PS reconstructions.
The Lorentz reconstruction 
algorithm, although only 
looking at the total 
$Q_{\rm ext}^{({\rm given})}
(\tilde\nu)$ 
of the 
layered sphere 
(single-element detector), 
correctly identifies 
PMMA in the core and 
PS in the shell. 
In addition, 
although the initial guesses 
of the complex refractive 
indexes 
are random (green lines), 
the reconstructions (magenta 
lines) of real and imaginary 
parts are close to their 
experimental values 
(blue lines). 
This shows that space-resolved 
chemistry (in this case 
for a sphere with two layers)
is possible with a single-element detector recording only 
radiation emitted 
in forward direction.}
\label{fig:fig4}
\end{figure}

In Section~\ref{HOM-SPH} 
we concentrated on 
exploring the power of 
our Lorentzian method 
in the well-defined case 
of homogeneous spheres. 
We now ask 
the question 
whether our 
Lorentzian algorithm 
is powerful enough to 
obtain the space-resolved 
complex refractive index, 
i.e., via the 
imaginary part of 
the complex refractive 
index (absorption), 
space-resolved chemistry. 
To answer this question, we
focus on a 
stratified sphere 
with two layers. 
We fill the core of 
the layered sphere 
with PMMA, the shell 
with PS, and compute 
$Q_{\rm ext}^{({\rm given})}
(\tilde\nu)$ 
on the basis of 
(\ref{L-sum}) with 
$M=2$. 
The outer radius 
of the sphere 
is $10\mu$m; 
the inner radius, 
i.e., the core radius, 
is $8\mu$m. 
For the PMMA 
refractive index 
in the core 
we used experimental 
PMMA data. 
For the PS 
refractive index 
in the shell, 
we used the known 
imaginary part of the 
refractive index and 
computed the real part 
using the 
Kramers-Kronig transformation. 
We emphasize that the 
use of the Kramers-Kronig 
transformation in this case 
has nothing to do with 
our reconstruction algorithm 
(which never uses the 
Kramers-Kronig transformation), 
and is used only to generate 
the real part of the 
PS refractive index. 
We could have, as well, 
used a complete set of 
PS experimental data 
for real and imaginary 
parts of 
the refractive 
index, avoiding 
the Kramers-Kronig transformation 
altogether, but we were unable 
to locate a suitable, published 
set. 
In any case, 
constructing 
$Q_{\rm ext}^{({\rm given})}
(\tilde\nu)$
on the basis of this 
input, we 
obtained the reconstructions 
of the complex indexes 
of refraction in the core 
and in the shell 
as shown in 
Fig.~\ref{fig:fig4}. 

The result of our 
Lorentzian reconstruction 
of the space-dependent 
complex refractive 
index is shown 
in 
Fig.~\ref{fig:fig4}. 
We see that, not only 
does our Lorentzian 
algorithm, based only 
on the extinction efficiency, 
correctly identify the 
two substances and their 
spatial locations 
(PMMA in the core and 
PS in the shell), the 
reconstructions 
themselves (magenta lines) 
are close to their 
actual values (blue lines). 
This is encouraging, the more 
that, again, we started from 
completely unbiased, 
random initial conditions 
(green lines). 
This demonstrates that 
our Lorentzian algorithm 
is capable of 
revealing space-resolved 
chemistry based only 
on a single-element detector. 
We mention that we also 
used our Lorentzian method to 
discover the core and 
sphere radii, 
assumed unknown. 
In this case we add the 
core and sphere radii 
as additional parameters 
to our Lorentzian method. 
We found that in this case 
our Lorentzian method is 
not only able to reconstruct 
the complex indexes of 
refraction, but simultaneously 
also determines the core 
and sphere radii. 
As a result of this 
type-IV simulation, we 
find that 
in this case, too, we can 
reconstruct the complex 
refractive indexes, 
space- and 
wavenumber-dependent, 
simultaneously 
with determining sphere size and 
information on internal 
structure (in this case 
the core radius, which 
may stand for the nucleus 
of a biological cell). 

%
%
%
 
\section{Discussion} 
\label{DISC} 
In contrast to other 
spectroscopic methods, 
such as, e.g., Raman 
spectroscopy, the 
IR power in IR 
spectroscopy is low  
and thus does not harm 
the cells. 
In addition, since IR 
spectroscopy does not 
require staining, 
IR spectroscopy does 
not alter the cells or 
the cell chemistry. 
Thus, IR spectroscopy 
enables in-vivo investigations 
of cells and tissues. 
In their natural state, 
biological cells have 
a 3D structure, which, 
combined with the fact 
that cells and tissues 
have sizes that are 
of the order of the
IR wavelength, leads 
to large scattering 
contributions that 
need to be removed 
to obtain pure absorbance 
spectra that only then 
can be interpreted 
as to their chemical 
information content. 
Our new Lorentzian method, 
compact and powerful,  
accomplishes 
just that, and, 
as demonstrated in  
Section~\ref{RESULTS}, 
can even 
be combined with 
baseline, tilt, 
curvature, and 
scaling 
corrections of raw spectra. 
The method is also suitable 
for extracting space-resolved 
chemistry as demonstrated 
in Section~\ref{LAY-SPH}
with the help of 
a layered sphere. 
 
Our results also show that 
focal plane arrays (FPAs) 
may not be 
necessary to obtain 
space-resolved chemistry. 
For instance, 
as shown  
for the layered sphere, 
a single-element 
detector suffices. 
FPAs may still be needed 
if the scatterer has 
a more detailed 
shape or 
internal structure. 
However, rotating 
the sample, 
or using different 
illumination directions, 
akin to 
X-ray tomography, 
we may still get away 
with only a single-element 
detector. 
 
Concerning stability and 
reliability of our Lorentzian 
method, 
we performed several checks. 
(a) We perturbed the simulated 
$Q_{\rm ext}$ 
spectrum with 
up to 10\% of noise and found 
that reconstruction was still 
possible. 
(b) We always start our reconstructions 
with random inputs for 
peak widths, peak heights, and 
peak locations. We always find 
convergence to the correct 
pure absorbance spectrum. 
This demonstrates that 
our method is robust 
and does not require fine-tuning 
of input parameters. 

With the advent of new 
IR sources, such as 
tunable 
quantum cascade lasers, 
IR spectroscopy has 
recently received 
a rejuvenating boost 
that is ideally suited 
for the application 
of our method, which 
works well with 
coherent illumination 
of samples. 
We mention that 
in the case of coherent IR 
sources, optical 
effects, such as 
interference and 
scattering are 
more important 
and more controlled 
as compared with 
thermal IR sources 
(e.g., GLOBAR) or 
(partially coherent) 
synchrotron sources. 
Thus, 
our Lorentzian method is 
ideally suited for 
obtaining pure absorbance 
spectra in combination 
with coherent laser sources. 
We also mention that 
lasers are cheaper 
and more accessible 
than synchrotron sources, 
which makes our method 
ideally suited for 
laboratory applications. 
%

\section{Conclusion} 
\label{CONC}
In this paper
we have shown that it is 
possible to reconstruct 
the spatially
resolved, 
wavenumber-dependent 
complex 
index of refraction 
(the spatially resolved 
chemistry) of a sample 
on the basis of 
extinction spectra 
which require only 
a single-element 
detector. 
The method is robust 
and reliable. 
It can be extended 
in several directions. 
(a) How 
many different layers 
of a stratified sphere 
can we resolve with a 
single-element detector? 
(b) Extension 
of the method to 
spheres with 
internal structure that 
does not have rotational 
symmetry.  
(c) Extension to 
3D shapes with 3D internal 
structure. 
The most 
important question 
to answer is 
the following: 
Up to what point is a 
single-element detector 
enough, and at what point 
is it more advantageous to 
work with an FPA? 
These topics 
define exciting 
directions for future research. 
In addition, 
our method may find practical 
applications in biophysics, 
including in medical 
diagnostics, medical pathology, 
and 
single-cell spectroscopy. 
 
\section{Acknowledgments}
\label{ACK}
We thank the experimental 
team at the Norwegian 
University of 
Life Sciences 
for the raw data of 
the 
experimental PMMA
$Q_{\rm ext}$ spectrum 
published in 
\cite{blumel2016PMMA} 
which was used for 
the reconstructions 
shown in 
Fig.~\ref{fig:fig3}. 
 
\section{Author contributions} 
RB suggested the 
method and 
supervised the project. 
KdS coded and performed 
all computer simulations. 
All authors contributed 
to the writing of the paper. 
 
\bibliographystyle{alpha}
\bibliography{sample}
\end{document}